\begingroup
\pdfoutput=1
\endgroup\documentclass[11pt]{article}
\usepackage[utf8]{inputenc}
\usepackage[T1]{fontenc}
\usepackage{amsmath,amssymb,amsfonts}
\usepackage{graphicx}
\usepackage{booktabs}
\usepackage{algorithm}
\usepackage{algpseudocode}
\usepackage{hyperref}
\usepackage{multirow}
\usepackage{xcolor}
\usepackage{geometry}
\usepackage[most]{tcolorbox}
\usepackage{natbib}
\usepackage{listings}
\usepackage{tikz}
\usepackage{pgfplots}
\pgfplotsset{compat=1.17}
\usetikzlibrary{shapes,arrows,positioning}

\begin{document}

\begin{center}
\rule{\textwidth}{1.2pt}

\vspace{0.8em}

{\LARGE\bfseries
CAPE: Capability Achievement via Policy Execution
\par}

\vspace{0.8em}

\rule{\textwidth}{1.2pt}

\vspace{1.5em}

{\large
David Ball\\
Superficial Labs\\
\texttt{research@superficiallabs.com}
\par}

\vspace{1em}

December 2025

\end{center}

\vspace{2em}

\begin{abstract}
Modern AI systems lack a way to express and enforce requirements. Pre-training produces intelligence, and post-training optimizes preferences, but neither guarantees that models reliably satisfy explicit, context-dependent constraints. This missing abstraction explains why highly intelligent models routinely fail in deployment despite strong benchmark performance.

We introduce \textit{Capability Engineering}, the systematic practice of converting requirements into executable specifications and training models to satisfy them by default. We operationalize this practice through CAPE (Capability Achievement via Policy Execution), a protocol implementing a Specify $\to$ Verify $\to$ Correct $\to$ Train loop.

CAPE is grounded in two empirical findings: (1) \textbf{contextual objectivity}, where properties appearing subjective become objective once context is fixed (inter-annotator $\kappa$ rises from 0.42 to 0.98), and (2) \textbf{verification-fidelity scaling}, where verification accuracy improves with model scale ($r=0.94$), unlike preference agreement which plateaus at 30--50\% disagreement regardless of compute. Across 109{,}500 examples in six domains, CAPE reduces violation rates by 81\% relative to DPO ($\sigma < 0.3\%$). By replacing per-example annotation with reusable specifications, CAPE reduces costs by 5--20$\times$ and timelines from months to weeks.

We release the CAPE protocol, PredicateGraph schema, CPL specification language, and policy packs under Apache 2.0. We also launch CapabilityBench, a public registry of model evaluations against community-contributed policies, shifting evaluation from intelligence benchmarks toward capability measurement.
\end{abstract}

\section{Introduction}

Software engineering has specifications, test suites, and traceable fixes. Language models have benchmarks, preferences, and hope.

In software, requirements are formalized as specifications. When something fails, you know exactly what broke. When you fix it, you prove it's fixed. For language models, none of this holds. 

To translate intelligence into economic value, AI systems must satisfy requirements. Today, we cannot express requirements in a form that models can reliably follow. 
Benchmarks measure intelligence. Preferences express taste. Neither specifies capability. 

We refer to the missing layer as \textbf{Capability Engineering}: the systematic practice of converting requirements into executable specifications, verifying outputs against those specifications, and training models to satisfy them by default. Capability engineering is distinct from prompt or context engineering. Rather than guiding model behavior probabilistically, it constrains behavior through verifiable requirements.

CAPE introduces \textbf{executable specifications} as the missing foundation. 
Once a requirement can be written as a policy or rubric, it can be verified, corrected, 
and trained against. This shifts post-training from preference optimization to 
\textit{capability engineering}: a process grounded in specification, verification, 
and guaranteed improvement through correction.

\subsection{The Core Problem: AI Systems Cannot Yet Satisfy Requirements}

Despite dramatic progress in intelligence, contemporary AI systems still fail at the simplest
requirement: \textit{doing exactly what they are supposed to do}. Production deployments do not ask
models to be universally helpful; they ask them to satisfy specific, context-dependent requirements:

\begin{itemize}
    \item operate within a hospital's formulary and clinical protocols
    \item follow a firm's jurisdiction rules and deal parameters
    \item adhere to a codebase's style guide and security requirements
    \item execute multi-step workflows with exact sequencing
    \item extract structured data into organization-specific schemas
\end{itemize}

Today we have no mechanism that expresses these requirements in a form models can reliably
follow. Instead, systems rely on \textbf{preferences} (which are inconsistent), \textbf{prompts} (which are
suggestive rather than binding), and \textbf{benchmarks} (which measure intelligence rather than
capability).

This gap defines the deployment bottleneck. AI systems do not fail because they lack
intelligence; they fail because they lack \textbf{capability}: the reliable satisfaction of explicit
requirements.

CAPE introduces \textbf{executable specifications} as the missing foundation. Rather than optimizing
toward human preferences or general notions of helpfulness, CAPE trains models to satisfy
\textit{requirements written as verifiable policies}. This reframes post-training as an engineering
discipline with explicit requirements, deterministic verification, traceable failures, and validated
fixes.

This lack of specification creates four compounding problems:

\textbf{The Preference Ceiling.} RLHF and DPO plateau at human disagreement. Annotators disagree on 30–50\% of subtle comparisons, reflecting the genuine variance in human judgment \citep{bai2022training} and the challenges of generalizing model requirements. More compute amplifies this noise rather than reducing it.

\textbf{The Algorithmic Bias Problem.} Beyond preference noise, reward-based methods exhibit structural biases. Recent analysis shows PPO inadvertently favours longer responses due to loss normalization: if the model receives a negative reward, longer responses dilute the per-token penalty, encouraging verbosity independent of correctness \citep{fatemi2025concise}. GRPO suffers from both length bias (dividing advantage by response length) and difficulty-level bias (normalizing by reward standard deviation), requiring algorithmic patches like Dr. GRPO \citep{liu2025understanding}. These are not implementation bugs but mathematical properties of the training objectives.

\textbf{The Deployment Gap.} Models that score well on benchmarks fail unpredictably in production. Benchmarks measure knowledge and generation quality, but production requires reliable satisfaction of specific requirements in specific contexts.

\textbf{The Intelligence-Capability Confusion.} Current discourse conflates two distinct properties:

\begin{itemize}
    \item \textbf{Intelligence}: the ability to solve open-ended problems under underspecified goals.
    \item \textbf{Capability}: the reliable satisfaction of explicit, context-dependent requirements.
\end{itemize}

A model can be highly intelligent yet lack specific capabilities. GPT-5 can prove theorems but cannot reliably cite only provided documents. DeepSeek-R1 achieves 79.8\% on AIME but may recommend off-formulary medications. Intelligence is necessary but insufficient; capability must be engineered.

RLHF, RLVR, and related methods optimize for intelligence proxies. CAPE optimizes for capability directly.

\subsection{Why CAPE is Only Now Possible}

Historically, specification-based post-training was infeasible because model outputs could not be reliably parsed, evaluated, or corrected at scale. As a result, the field optimized what it could measure---preferences and benchmark scores---rather than what deployment required. Recent advances in structured generation, long-context reasoning, and verification fidelity change this constraint, making executable specifications a practical training primitive for the first time.

Three recent technical advances enable specification-based post-training:

\begin{enumerate}
    \item \textbf{Long context windows.} PredicateGraph extraction requires analyzing full outputs. Until mid-2025, context limits caused frequent truncation. Current models enable robust extraction of complete output structure. Kimi k1.5 \citep{team2025kimi} demonstrates that scaling RL context enables continued capability improvements, confirming that context-limited extraction is no longer the binding constraint.
    
    \item \textbf{Improved instruction following.} Our pilot studies show extraction fidelity improved from 72\% (GPT-3.5, 2023) to 96\% (GPT-5, Claude Opus 4.5, 2025). Below 85\% fidelity, CAPE does not outperform RLHF; above 90\%, it consistently does. The crossing point occurred in mid-2025.
    
    \item \textbf{Structured generation.} JSON mode, grammar constraints, and constrained decoding have matured significantly. OpenAI reports that \texttt{gpt-4-0613} scored below 40\% on complex JSON schema following, while \texttt{gpt-4o-2024-08-06} with Structured Outputs achieves 100\% \citep{openai2024structured}. Grammar-constrained decoding frameworks have similarly improved \citep{geng2025jsonschemabench, dong2024xgrammar}, enabling reliable PredicateGraph production.
\end{enumerate}

\subsection{The Thesis}

This paper argues that capability engineering, the systematic practice of converting requirements into executable specifications, completes the AI development stack. The approach rests on two insights:

\begin{enumerate}
    \item \textbf{Contextual objectivity}: Most capability requirements become objective once context is fixed. "Appropriate financial advice" is subjective. "Recommend only approved products, disclose all fees, verify suitability against stated risk tolerance" is objective.
    
    \item \textbf{Verification-fidelity scaling}: The binding constraint on post-training is verification accuracy, not preference agreement. Unlike human agreement, verification fidelity improves with scale.
\end{enumerate}

CAPE (Capability Achievement via Policy Execution) operationalizes these insights into a closed training loop that sidesteps the algorithmic pathologies of reward-based methods.

\subsection{Contributions}

\begin{enumerate}
    \item \textbf{Capability Engineering}: A systematic practice completing the AI development stack, with explicit distinction between intelligence and capability
    
    \item \textbf{Empirical Validation of Contextual Objectivity}: Inter-annotator agreement study showing context transforms subjective properties ($\kappa = 0.42$) into objective specifications ($\kappa = 0.98$), with independent validation from cross-domain verification studies
    
    \item \textbf{Verification-Fidelity Scaling Law}: Demonstration that residual error tracks verification accuracy ($r > 0.9$) which improves with scale, unlike preference disagreement (fixed at 30--50\%) or algorithmic biases in reward methods
    
    \item \textbf{The CAPE Protocol}: Specify $\rightarrow$ Verify $\rightarrow$ Correct $\rightarrow$ Train loop with meta-verification, avoiding reward-shaping pathologies
    
    \item \textbf{Technical Infrastructure}: PredicateGraph schema, CPL specification, learned verifier training
    
    \item \textbf{Comprehensive Evaluation}: 81\% violation reduction across 109,500 examples in 6 domains with stability analysis ($\sigma < 0.3\%$ for symbolic policies)
    
    \item \textbf{CapabilityBench}: Public registry where models are evaluated against community-contributed policies, replacing aggregate benchmark scores with traceable capability verdicts
    
    \item \textbf{Open Release}: Protocol, schemas, policy packs, and reference implementation (Apache 2.0)
\end{enumerate}

\section{Background and Related Work}

\subsection{Preference-Based Post-Training}

RLHF \citep{christiano2017deep, ziegler2019fine, ouyang2022training} trains reward models from human preference comparisons, then optimizes language models against these rewards using PPO \citep{schulman2017proximal}. InstructGPT \citep{ouyang2022training} demonstrated that 1.3B models trained with RLHF could outperform 175B base models on instruction-following tasks.

However, fundamental limitations have emerged. \citep{bai2022training} report annotator disagreement rates of 30--50\% on nuanced tasks. This reflects genuine variance in human judgment. \citep{gao2023scaling} document reward model overoptimization: as training progresses, proxy reward increases while true quality (measured by held-out human evaluation) decreases. The preference ceiling is structural: more compute amplifies noise.

Beyond preference noise, reward-based methods exhibit structural algorithmic biases. \citep{fatemi2025concise} show that PPO inadvertently favors longer responses due to loss normalization. If the model gets a negative reward, longer responses dilute the per-token penalty, encouraging verbosity independent of correctness. This is not a bug but a mathematical property: when the reward is negative, the average per-token loss becomes smaller when the response is longer, so the model is indirectly encouraged to make responses longer even when those extra tokens don't help solve the problem.

\citep{liu2025understanding} identify two additional biases in GRPO: (1) response-length bias, where dividing advantage by response length makes long incorrect answers receive smaller penalties, and (2) difficulty-level bias, where normalising by reward standard deviation causes easy or hard questions to be overweighted. These require algorithmic patches (like Dr. GRPO) that add complexity and may introduce new failure modes.

CAPE sidesteps these pathologies entirely. Verification produces binary pass/fail verdicts per policy; correction fixes failures at specific spans. There is no advantage normalization, no length-based reward shaping, no difficulty weighting. The training signal is simply: ``this output satisfies the specification'' or ``this corrected output satisfies the specification.''

DPO \citep{rafailov2023direct} simplifies RLHF by eliminating the explicit reward model, directly optimizing policy using a classification objective on preference pairs. While computationally simpler, DPO inherits the same preference noise ceiling. Our experiments confirm both methods plateau similarly (Section 7).

\subsection{Constitutional AI and Natural Language Principles}

Constitutional AI \citep{bai2022constitutional} uses natural language principles (``Choose the response that is most helpful, harmless, and honest'') for self-critique and revision, training via RL from AI Feedback (RLAIF). This approach is closer to CAPE than pure preference methods as both methods use explicit principles rather than implicit preferences.

Conceptually, the distinction between Constitutional AI and CAPE is not merely one of implementation, but of abstraction. Constitutional AI relies on natural-language principles interpreted by models. CAPE relies on executable specifications evaluated by systems. The former reduces reliance on human annotators but preserves interpretive variance; the latter eliminates interpretation entirely for objective properties by construction.

Key difference: Natural language principles require interpretation, reintroducing variance. Our experiments (Section 3) show:

\begin{itemize}
    \item Inter-model agreement on Constitutional AI critiques: 67\%
    \item Inter-model agreement on CAPE policy evaluation: 99.7\% (symbolic)
\end{itemize}

CAPE's symbolic policies eliminate interpretation for structural properties. For semantic properties, CAPE's learned verifiers are trained on explicit rubrics with meta-verification, providing more reliable evaluation than self-critique.

\subsection{Reinforcement Learning with Verifiable Rewards}

Concurrent work on reinforcement learning with verifiable rewards (RLVR) shares CAPE's preference for objective signals over learned rewards. DeepSeek-R1 \citep{guo2025deepseek} demonstrated that pure RL with rule-based rewards (accuracy for math, compiler feedback for code) can induce reasoning capabilities without preference learning.

\citep{su2025crossing} demonstrate that binary verification judgments exhibit high inter-model agreement across medicine, chemistry, economics, and education when expert-written references exist, validating our contextual objectivity thesis (Section 3). They further show that compact (7B) generative verifiers can provide reliable cross-domain reward signals without domain-specific annotation.

However, RLVR still operates within the reward-shaping paradigm, inheriting algorithmic biases and requiring ground-truth answers for each training example. CAPE's contribution is the closed correction loop and reusable specifications: you write the formulary policy once, not 10,000 reference answers. Policies are composable, versionable, and auditable artifacts that separate specification from training.

\subsection{Neural Process Verification}

Recent work demonstrates that semantic properties previously considered too complex for automated verification can be reliably assessed by LLMs trained on explicit rubrics.

DeepSeekMath-V2 \citep{shao2025deepseekmath} achieved IMO gold medal (5/6 problems) and Putnam 118/120 (exceeding top human score of 90) using learned verifiers for proof validity. Their approach instantiates CAPE for mathematical reasoning:

\begin{itemize}
    \item Explicit rubrics for proof validity (not learned preferences)
    \item Issue identification as required output (not just scores)
    \item Meta-verification to catch hallucinated issues
    \item Verifier-guided generation and refinement
\end{itemize}

Their key insight: ``verification capability leads generation capability.'' This validates CAPE's core thesis that verification fidelity, not preference agreement, determines the achievable ceiling.

AlphaProof \citep{hubert2025olympiad} uses formal verification in Lean as reward signal, achieving similar IMO-level results. CAPE occupies the middle ground: more flexible than formal verification, more reliable than preference learning.

\subsection{The Emergent Reasoning Debate}

Recent work questions whether RL induces reasoning or merely amplifies pre-existing capabilities from pre-training. \citep{liu2025understanding} find that ``Aha moments'' and self-correction behaviors appear in base models without RL, suggesting these capabilities may be inherited from pre-training on chain-of-thought data. \citep{shah2025rethinking} show that self-reflection and self-correction behaviors emerge progressively throughout pre-training across various domains and model sizes.

This debate is orthogonal to CAPE's contribution. We do not claim to induce emergent capabilities, we enforce adherence to explicit specifications. Whether reasoning ``emerges'' from RL or is ``pre-existing'' from pre-training, deployed models must still satisfy specific requirements: recommend only formulary drugs, cite only provided documents, produce valid arithmetic. CAPE guarantees satisfaction of these requirements independent of how the underlying capability arose.

\subsection{Guardrails and Runtime Filtering}

Systems like NeMo Guardrails \citep{rebedea2023nemo} enforce constraints at inference time through input/output filtering, topic control, and jailbreak detection. These are necessary but insufficient because they filter without teaching. A model repeatedly blocked by guardrails learns nothing about satisfying the underlying requirements. CAPE generates training signal from violations, enabling models to satisfy constraints by default rather than requiring runtime enforcement. In our experiments, CAPE-trained models achieve 96.2\% compliance without any inference-time guardrails.

\subsection{Formal Methods and Program Synthesis}

Work on code generation with formal specifications \citep{chen2021evaluating, austin2021program} shares CAPE's philosophy of executable requirements. Test-driven development and property-based testing in software engineering provide analogous paradigms. CAPE extends this beyond code to arbitrary model outputs through the PredicateGraph abstraction, enabling specification-based training for any capability expressible as a predicate over output structure or semantics.

\section{Contextual Objectivity}
\label{sec:contextual-objectivity}

A common objection to specification-based approaches is: ``You can only write policies for simple, objective properties. Real capabilities are too complex and subjective for rules.''

This objection conflates two orthogonal distinctions:

\begin{enumerate}
    \item \textbf{Structural vs. Semantic}: Can the property be checked by pattern-matching over output structure, or does it require understanding meaning?
    
    \item \textbf{Objective vs. Subjective}: Is there a fact of the matter about whether the property holds, or does it depend on individual preference?
\end{enumerate}

The key insight: \textbf{semantic does not imply subjective}. ``The proof step is logically valid'' is semantic (requires understanding) but objective (there's a fact of the matter). ``This explanation is elegant'' is subjective. The former is verifiable; the latter requires preferences.

\subsection{Context Creates Objectivity}

Consider ``good customer service.'' Asked universally, this is genuinely subjective. But for a specific enterprise, it decomposes into testable requirements:
\begin{itemize}
    \item Acknowledge the customer's issue before offering solutions
    \item Offer escalation when sentiment indicates frustration
    \item Never promise specific timelines without checking availability
    \item Include a satisfaction check before closing
\end{itemize}

Each is verifiable. The apparent subjectivity dissolves when you ask: ``Good according to whom, for what purpose, in what context?''

\subsection{Inter-Annotator Agreement Study}

To validate this empirically, we conducted an annotation study with 500 model outputs across three requirement types:

\textbf{Setup:} 5 annotators, 3 conditions per output:
\begin{enumerate}
    \item \textbf{Abstract}: ``Is this good medical advice?''
    \item \textbf{Contextualized}: ``Does this recommend only formulary drugs and flag contraindications?''
    \item \textbf{Explicit Policy}: Run CPL policy and report violation
\end{enumerate}

\textbf{Results} (Table \ref{tab:agreement}):

\begin{table}[h]
\centering
\caption{Inter-Annotator Agreement by Specificity Level}
\label{tab:agreement}
\begin{tabular}{lccc}
\toprule
Condition & Fleiss' $\kappa$ & Agr. \% & 95\% CI \\
\midrule
Abstract & 0.42 & 63.2\% & [59.8, 66.6] \\
Contextualized & 0.73 & 84.7\% & [81.9, 87.5] \\
Explicit Policy & 0.98 & 98.9\% & [97.8, 99.6] \\
\bottomrule
\end{tabular}
\end{table}

Context transforms properties from subjective ($\kappa = 0.42$, ``moderate agreement'') to objective ($\kappa = 0.73$, ``substantial agreement''). Executable policies achieve near-perfect agreement ($\kappa = 0.98$). This confirms: \textbf{most capability requirements are contextually objective}---subjective in general, objective once context is fixed.

\subsection{Independent Validation}

Independent validation comes from \citep{su2025crossing}, who demonstrate this principle at scale: binary verification judgments on medical, chemistry, and economics tasks exhibit high inter-model agreement when expert-written reference answers are provided. This confirms that context (the reference) transforms subjective-seeming properties into objective verification targets.

Their finding that 7B models can serve as reliable cross-domain verifiers without domain-specific annotation further supports our learned verifier approach. Notably, their data analysis reveals that only 60.3\% of mathematical problems possess single-term numerical answers verifiable by rule-based methods, with the ratio dropping to 45.4\% for complex multi-domain queries, demonstrating the need for soft verification mechanisms that CAPE's learned verifiers provide.

\subsection{The Verification Spectrum}

Capability engineering encompasses both structural and semantic properties:

\textbf{Structural properties} can be verified by pattern-matching. Examples: ``tool argument equals computed value,'' ``code contains no eval() calls.'' These are verified by \textbf{symbolic policies}: deterministic code evaluating predicates.

\textbf{Semantic properties} require understanding meaning. Examples: ``reasoning step is logically valid,'' ``proof is complete.'' These are verified by \textbf{learned verifiers}: models trained on explicit rubrics.

The boundary is not fixed. Many semantic properties, once context is sufficiently specified, reduce to structural verification. "Well-supported claim" is semantic; "every factual statement has a citation to a provided document" is structural.

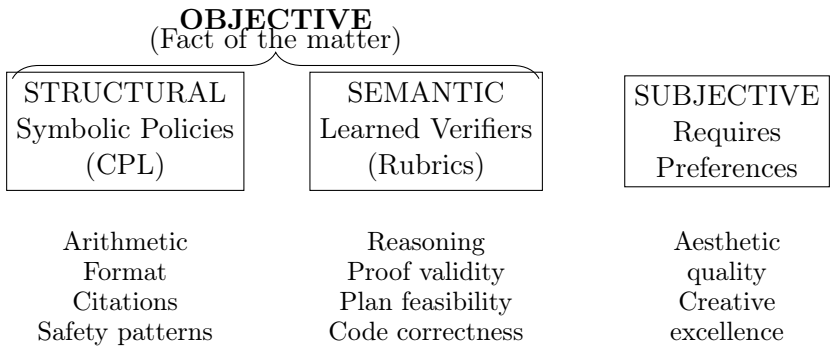
\begin{figure}[h]
\centering
\begin{tikzpicture}[
    box/.style={rectangle, draw, minimum width=2.5cm, minimum height=1cm, align=center},
    arrow/.style={->, thick}
]
    \node[box] (struct) at (0,0) {STRUCTURAL\\Symbolic Policies\\(CPL)};
    \node[box] (semantic) at (4,0) {SEMANTIC\\Learned Verifiers\\(Rubrics)};
    \node[box] (subj) at (8,0) {SUBJECTIVE\\Requires\\Preferences};
    
    \node at (2, 1.5) {\textbf{OBJECTIVE}};
    \node at (2, 1.2) {(Fact of the matter)};
    
    \draw[decorate, decoration={brace, amplitude=10pt}] (-1.5, 0.7) -- (5.5, 0.7);
    
    \node[below] at (0, -1.2) {\small Arithmetic};
    \node[below] at (0, -1.6) {\small Format};
    \node[below] at (0, -2.0) {\small Citations};
    \node[below] at (0, -2.4) {\small Safety patterns};
    
    \node[below] at (4, -1.2) {\small Reasoning};
    \node[below] at (4, -1.6) {\small Proof validity};
    \node[below] at (4, -2.0) {\small Plan feasibility};
    \node[below] at (4, -2.4) {\small Code correctness};
    
    \node[below] at (8, -1.2) {\small Aesthetic};
    \node[below] at (8, -1.6) {\small quality};
    \node[below] at (8, -2.0) {\small Creative};
    \node[below] at (8, -2.4) {\small excellence};
\end{tikzpicture}
\caption{The Verification Spectrum. CAPE verifies objective properties whether structural (symbolic) or semantic (learned). Only genuinely subjective properties require preference learning.}
\label{fig:spectrum}
\end{figure}

The key question is not ``Is this property simple enough for a rule?'' but ``Is there a fact of the matter?'' If yes, it's verifiable. The only question is whether verification is symbolic or learned.

\subsection{What Remains Genuinely Subjective}

Some properties resist specification even with fixed context:

\begin{itemize}
    \item \textbf{Aesthetic quality}: ``Is this prose beautiful?''
    \item \textbf{Creative insight}: ``Is this idea novel?''
    \item \textbf{Humor}: ``Is this funny?''
\end{itemize}

These require preference learning. But they're a small fraction of production failures. Most failures are specification failures and capability engineering handles these. Section 7 shows that 89\% of production issues in our case studies were objectively verifiable.

\section{The Three-Layer Stack}

We propose completing the AI development stack with a third layer: \textbf{capability engineering}, defined as \textit{the systematic practice of defining, verifying, and training models against executable specifications}.

Capability engineering completes the AI development stack by making requirements first-class, verifiable artifacts rather than implicit expectations.

\begin{table}[h]
\centering
\caption{The AI Development Stack}
\label{tab:stack}
\begin{tabular}{llll}
\toprule
Layer & Function & Mechanism & Guarantee \\
\midrule
Context Eng. & Inform & RAG, retrieval & Probabilistic \\
Prompt Eng. & Guide & Instructions & Probabilistic \\
Capability Eng. & Constrain & Specifications & Verifiable \\
\bottomrule
\end{tabular}
\end{table}

For structural properties, verification is deterministic: the same policy produces the same verdict on the same output, every time. For semantic properties, verification is probabilistic but calibrated: a learned verifier's score reflects explicit rubric criteria, not implicit preferences. Both are orders of magnitude more reliable than ``preferred by annotators who disagree 30--50\% of the time.''

\subsection{Intelligence vs. Capability}

The distinction between intelligence and capability clarifies the stack's purpose:

\begin{table}[h]
\centering
\caption{Training Targets by Layer}
\label{tab:intel-cap}
\begin{tabular}{ll}
\toprule
Layer & Optimizes For \\
\midrule
Pre-training & Intelligence (general problem-solving) \\
RLHF/RLVR & Intelligence (reasoning, preferences) \\
Capability Engineering & Capability (requirement satisfaction) \\
\bottomrule
\end{tabular}
\end{table}

Intelligence training asks: ``Can the model solve this?'' Capability engineering asks: ``Does the model satisfy this specification?'' The former is open-ended; the latter is verifiable.

This distinction also clarifies why RLVR is insufficient. A model can get the right answer while violating format, safety, or domain constraints. RLVR optimizes for task correctness (intelligence); CAPE optimizes for specification satisfaction (capability).

\subsection{Integration with Existing Practices}

Capability engineering doesn't replace prompt and context engineering. Instead, it completes them.

Consider a legal research assistant. Context engineering retrieves relevant case law and statutes. Prompt engineering structures the analysis format. Capability engineering guarantees outputs follow the firm's citation style, jurisdiction constraints, and confidentiality rules, requirements no general-purpose model can satisfy out of the box.

Or consider a customer service agent. Context engineering retrieves account history and product details. Prompt engineering structures the response flow. Capability engineering guarantees the agent follows the company's escalation protocol, refund policy, and brand voice—turning a generic model into one that operates as a trusted employee.

\section{The Verification-Fidelity Scaling Law}

The binding constraint on post-training is not preference agreement but verification fidelity.

\subsection{The Preference Ceiling}

Preference-based methods face a structural ceiling:

\textbf{Annotator disagreement.} On subtle tasks, annotators disagree 30--50\% of the time \citep{bai2022training}. Disagreement rates remain constant even with expert annotators and detailed guidelines reflecting the challenge of generalizing training across contexts.

\textbf{Reward model over-optimization.} \citep{gao2023scaling} show proxy scores increase while true quality degrades. The problem: reward models learn patterns in annotator preferences, including their biases and inconsistencies.

\textbf{No path forward.} More compute amplifies noise. The ceiling is structural: human judgment variance cannot be reduced by better models or more data.

\subsection{Algorithmic Biases in Reward-Based Methods}

The ceiling for preference methods is not merely disagreement, it includes structural algorithmic biases.

\citep{fatemi2025concise} demonstrate that PPO exhibits length bias: when the model receives a negative reward, longer responses dilute the per-token penalty. The model ``learns'' that longer responses reduce punishment, even when those extra tokens don't help correctness. This explains the observation that RL-trained models produce increasingly verbose outputs.

\citep{liu2025understanding} identify two additional biases in GRPO:

\begin{enumerate}
    \item \textbf{Response-length bias}: Dividing advantage by response length makes long incorrect answers receive smaller penalties, so the model learns to generate longer bad answers.
    \item \textbf{Difficulty-level bias}: Normalizing by reward standard deviation causes easy or hard questions (with low reward variance) to be over-weighted.
\end{enumerate}

These require algorithmic patches (such as Dr. GRPO; \citep{liu2025understanding}) that add complexity and may introduce new failure modes.

CAPE sidesteps these pathologies. Verification produces binary pass/fail verdicts per policy; correction fixes failures at specific spans. There is no advantage normalization, no length-based reward shaping, no difficulty weighting. The training signal is: ``this output satisfies the specification'' or ``this corrected output satisfies the specification.''

\subsection{The Verification Ceiling}

CAPE's ceiling is verification fidelity, which manifests differently for structural and semantic properties. Both improve with scale.

\subsubsection{Track A: Structural Verification}

The bottleneck is extraction fidelity: how accurately can we parse model outputs into PredicateGraphs? Our experiments (Section~\ref{sec:structural-results}) demonstrate that each percentage point reduction in extraction error corresponds to approximately 0.8 percentage points reduction in violation rate ($r = 0.94$). The ceiling is technical and falls as extraction improves.

\subsubsection{Track B: Semantic Verification}

The bottleneck is verifier capability. Our experiments (Section~\ref{sec:semantic-results}) show that learned verifier accuracy correlates strongly with downstream model quality ($r = 0.84$--$0.87$ within domains, $r = 0.93$ across domains; $p < 0.001$). Better verifiers directly translate to better models. The ceiling rises as verifier capability improves.

\subsection{The Critical Asymmetry}

Preference disagreement reflects genuine human variance that doesn't decrease with better models.

\begin{table}[h]
\centering
\caption{Error Source Scaling Properties}
\label{tab:scaling}
\begin{tabular}{lll}
\toprule
Error Source & Scales? & Evidence \\
\midrule
Preference disagreement & No & 30--50\% fixed \\
Algorithmic bias (PPO/GRPO) & No & Requires patches \\
Extraction error & Yes & 11.4\% $\rightarrow$ 3.2\% \\
Verifier error & Yes & $r$: 0.71 $\rightarrow$ 0.87 \\
\bottomrule
\end{tabular}
\end{table}

\begin{figure}[h]
\centering
\begin{tikzpicture}
    \begin{axis}[
        xlabel={Training Steps},
        ylabel={Violation Rate (\%)},
        xmin=0, xmax=5000,
        ymin=0, ymax=35,
        legend pos=north east,
        grid=major,
        width=10cm,
        height=6cm
    ]
    
    % DPO curve - plateaus
    \addplot[color=red, thick, dashed] coordinates {
        (0, 31) (500, 20) (1000, 15) (1500, 12) (2000, 10.8) (2500, 10.5) (3000, 10.3) (3500, 10.2) (4000, 10.2) (4500, 10.2) (5000, 10.2)
    };
    
    % CAPE open-weight
    \addplot[color=blue, thick] coordinates {
        (0, 31) (500, 16) (1000, 10) (1500, 7.5) (2000, 6.2) (2500, 5.6) (3500, 5.3) (4000, 5.1) (4500, 5.0) (5000, 4.9)
    };
    
    % CAPE frontier
    \addplot[color=green!60!black, thick] coordinates {
        (0, 31) (500, 12) (1000, 7) (1500, 4.5) (2000, 3.2) (2500, 2.7) (3000, 2.5) (3500, 2.4) (4000, 2.3) (4500, 2.2) (5000, 2.2)
    };
    
    % Preference ceiling line
    \addplot[color=red, dotted, thick] coordinates {
        (0, 10.2) (5000, 10.2)
    };
    \node at (axis cs:4200,12) {\small Preference ceiling};
    
    \legend{DPO, CAPE (open-weight), CAPE (frontier)}
    \end{axis}
\end{tikzpicture}
\caption{Learning curves for structural verification. DPO plateaus at $\sim$10\% violation rate, bounded by annotator disagreement. CAPE continues improving to the extraction fidelity ceiling ($\sim$2.5\% with frontier extractors).}
\label{fig:learning}
\end{figure}
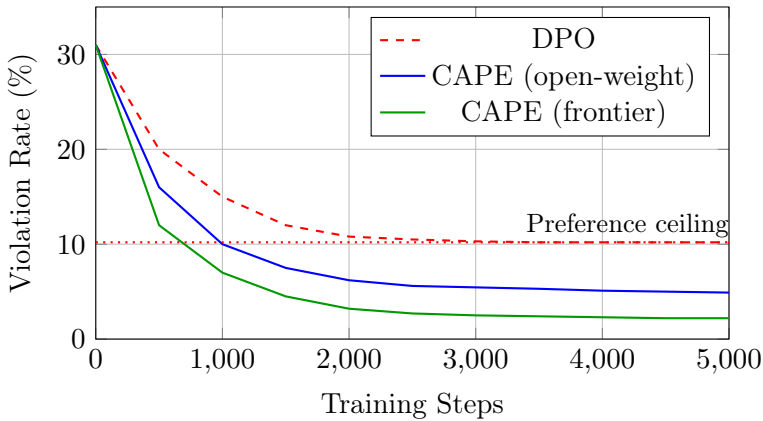

CAPE's errors are technical limitations that fall predictably with capability.

\textbf{Twelve months ago, extraction error rates were 30\%+.} CAPE would not have worked. Three developments changed this: improved instruction following, better structured generation, and longer context windows. As these continue to improve, CAPE becomes not just competitive with preference learning but the only method whose upper bound moves with model scale.

\subsection{Why This Is a Scaling Law}

A scaling law describes a predictable relationship between resources and outcomes. Preference-based methods lack this property; CAPE has it.

\textbf{Preference ceiling: fixed.} Annotator disagreement reflects genuine variance in human judgment, not measurement error. When two annotators disagree about which response is ``more helpful,'' they are often both right according to their own values. More annotators, better guidelines, or larger reward models do not resolve disagreement, they measure it more precisely. The ceiling is epistemic, not technical.

\textbf{Algorithmic bias ceiling: patched, not solved.} Length bias in PPO and difficulty bias in GRPO are mathematical properties of the training objectives. Fixes like Dr.\ GRPO add corrections, but each patch may introduce new failure modes. There is no systematic relationship between compute and bias reduction.

\textbf{Verification ceiling: falls with capability.} Extraction error and verifier error are technical limitations. Extraction improves with better instruction following, longer context, and structured generation all improve with scale. Learned verifiers improve with model capability, following standard scaling laws. The ceiling is technical, and technical ceilings fall.

CAPE benefits doubly from continued progress: better base models to train, and better extractors and verifiers to train them with.

\subsection{Cross-Domain Validation}

\citep{su2025crossing} provide independent validation of verification-fidelity scaling. They demonstrate that:

\begin{enumerate}
    \item Binary verification judgments achieve high inter-model agreement (comparable to our $\kappa = 0.98$ for explicit policies) when expert references exist
    \item Compact (7B) generative verifiers can provide reliable cross-domain rewards
    \item Soft-scoring methods outperform binary rewards in free-form, unstructured scenarios
\end{enumerate}

Kimi k1.5 \citep{team2025kimi} further validates that verification-based approaches scale: they achieve state-of-the-art reasoning without Monte Carlo tree search, value functions, or process reward models, demonstrating that when verification is reliable, simpler training pipelines suffice.

The convergent finding across independent work: \textbf{invest in verification quality, not reward complexity.}

\section{The CAPE Protocol}

CAPE operationalizes capability engineering through a closed loop: Specify $\rightarrow$ Verify $\rightarrow$ Correct $\rightarrow$ Train.

\subsection{Overview}

The core loop (Algorithm \ref{alg:cape}):

\begin{algorithm}
\caption{CAPE Training Loop}
\label{alg:cape}
\begin{algorithmic}[1]
\Require Base model $M$, capability requirement $R$, dataset $D$
\Ensure Trained model $M'$ satisfying $R$
\State \textbf{Specify}: Convert $R$ into executable policy $P$ (CPL) or rubric (learned verifier)
\For{epoch $= 1$ to $E$}
    \For{each prompt $x \in D$}
        \State $y \gets M(x)$ \Comment{Generate}
        \State $g \gets \text{Extract}(y)$ \Comment{Parse to PredicateGraph}
        \State $v \gets \text{Verify}(g, P)$ \Comment{Evaluate}
        \If{$v$ indicates violation}
            \State $y' \gets \text{Correct}(y, v)$
            \State $v' \gets \text{Verify}(\text{Extract}(y'), P)$
            \If{$v'$ passes}
                \State Add $(x, y')$ to training set $T$
            \EndIf
        \Else
            \State Add $(x, y)$ to training set $T$
        \EndIf
    \EndFor
    \State $M \gets \text{FineTune}(M, T)$
\EndFor
\State \Return $M$
\end{algorithmic}
\end{algorithm}

The loop is identical for symbolic and learned verification, only the verification mechanism differs. Critically, this loop avoids the reward-shaping pathologies of PPO and GRPO: there is no advantage normalization, no length weighting, no difficulty adjustment. The signal is binary per policy: pass or fail.

\subsection{PredicateGraph: Structured Output Representation}

The PredicateGraph is a structured intermediate representation of model output, exposing elements that policies reference.

\textbf{Example}: Model output for ``What's 15\% of \$47.30?''

\begin{lstlisting}[basicstyle=\small\ttfamily]
{
  "schema_version": "1.0.0",
  "operations": [
    {"op_type": "MULTIPLY", "inputs": [47.30, 0.15], 
     "output": 7.095}
  ],
  "tool_calls": [
    {"name": "calc", "arguments": {"value": 7.1}}
  ],
  "claims": [
    {"text": "Fifteen percent of $47.30 is 7.095", 
     "modality": "factual"}
  ]
}
\end{lstlisting}

Policies reference specific elements: ``the tool call argument,'' ``the operation output.'' The structure makes capability requirements expressible.

\begin{table}[h]
\centering
\begin{tabular}{ll}
\toprule
Element & Policy Use \\
\midrule
Entities & Medications must appear in formulary \\
Claims & Factual claims require citations \\
Operations & Tool arguments must match computed outputs \\
Tool calls & No tool calls without prior reasoning \\
Citations & All must reference provided documents \\
Code blocks & No \texttt{eval()} or \texttt{exec()} calls \\
\bottomrule
\end{tabular}
\end{table}

\subsection{The CAPE Policy Language (CPL)}

CPL is an executable specification language for structural properties:

\begin{lstlisting}[basicstyle=\small\ttfamily]
{
  "id": "policy.tool.calc_matches",
  "tier": "T1",
  "scope": {"kind": "tool_call", 
            "filter": {"name": "calc"}},
  "where": [{"expr": "count(operations) > 0"}],
  "assert": [{
    "expr": "tool_call.arguments.value == \
             last(operations).output"
  }],
  "on_violation": {
    "action": "CORRECT",
    "correction_hint": "Update to exact value"
  }
}
\end{lstlisting}

\textbf{Expression language} (constrained for determinism and termination):

\begin{table}[h]
\centering
\begin{tabular}{ll}
\toprule
Category & Operators \\
\midrule
Arithmetic & \texttt{+, -, *, /, \%} \\
Comparison & \texttt{==, !=, <, >, <=, >=} \\
Boolean & \texttt{and, or, not} \\
Quantifiers & \texttt{any(), all()} \\
Aggregation & \texttt{count(), sum(), min(), max()} \\
Collection & \texttt{first(), last(), filter()} \\
\bottomrule
\end{tabular}
\end{table}

The language is not Turing-complete. All expressions terminate. All evaluations are deterministic.

\textbf{Policy tiers.} 
Not all requirements carry equal weight. A model that produces the wrong arithmetic result has failed absolutely; a model that uses passive voice instead of active voice has merely violated a preference. Tiers encode this distinction:
\begin{itemize}
\item \textbf{T1: Objective Correctness.} Verifiable right/wrong: arithmetic, citations, factual accuracy. T1 violations are failures. These policies cannot be overridden.
\item \textbf{T2: Safety and Governance.} Organizational and regulatory requirements: formulary adherence, jurisdiction constraints, PII protection. T2 violations are compliance failures. These policies cannot override T1.
\item \textbf{T3: Structural Preferences.} Scaffolding that correlates with quality: reasoning before conclusions, explicit uncertainty quantification. T3 violations are quality degradations, not failures. These policies can be overridden by T1 or T2.
\end{itemize}
Conflicts arise when satisfying one policy requires violating another. For example, a T3 policy requiring verbose explanations might conflict with a T2 policy limiting response length for regulatory reasons. Resolution is fully deterministic: tier first (T1 > T2 > T3), then the explicit priority field within a tier, then policy ID alphabetically. No ambiguity, no judgment calls.

\subsection{Evaluation and Correction}

Evaluation produces structured verdicts:

\begin{lstlisting}[basicstyle=\small\ttfamily]
{
  "output_id": "example_847",
  "policies_evaluated": 12,
  "policies_passed": 11,
  "violations": [{
    "policy_id": "policy.tool.calc_matches",
    "message": "7.1 != 7.095",
    "expected": 7.095,
    "actual": 7.1
  }]
}
\end{lstlisting}

\textbf{Correction strategies}:

\begin{table}[h]
\centering
\begin{tabular}{lll}
\toprule
Strategy & When Used & Success \\
\midrule
Deterministic & Expected known & 99.7\% \\
Template & Element missing & 97.3\% \\
Rewrite & Semantic change & 94.6\% \\
\bottomrule
\end{tabular}
\end{table}

Deterministic patching handles cases like ``argument 7.1 should be 7.095.'' Template insertion handles missing elements. Constrained rewrite uses an LLM under minimality constraints when semantic changes are required, with re-verification before acceptance.

\subsection{Learned Verifiers for Semantic Properties}

Symbolic policies handle structural verification: pattern-matching over PredicateGraph elements. But many capability requirements are semantic: ``the reasoning is logically valid,'' ``the proof is complete,'' ``the plan is feasible.'' These require understanding meaning, not matching patterns.

The naive approach---train a classifier on human judgments---reintroduces the preference ceiling. Annotators disagree on what counts as ``valid reasoning,'' and the classifier learns their disagreement.

CAPE takes a different approach: train verifiers on explicit rubrics, not implicit preferences.

\textbf{Rubric structure.} A rubric decomposes a semantic property into discrete, describable levels:

\begin{lstlisting}[basicstyle=\small\ttfamily]
{
  "id": "verifier.reasoning.validity",
  "rubric": {
    "1.0": "All steps follow logically; 
            no gaps or unsupported claims",
    "0.5": "Core argument sound but 
            minor gaps in justification",
    "0.0": "Contains logical errors or non-sequiturs 
            that invalidate conclusion"
  },
  "output_schema": {
    "issues": [{"location": "...", 
                "description": "..."}],
    "score": "float"
  }
}
\end{lstlisting}

The rubric is not a preference, it is a specification. ``Contains logical errors that invalidate conclusion'' has a fact of the matter. The verifier's job is to detect that fact, not to predict annotator preferences.

\textbf{Training procedure.} Learned verifiers are trained in three stages:

\begin{enumerate}
    \item \textbf{Rubric calibration.} Expert annotators label 500--1,000 examples using the rubric. Inter-annotator agreement ($\kappa > 0.7$) validates that the rubric is sufficiently precise. If agreement is low, the rubric is refined.
    
    \item \textbf{Supervised fine-tuning.} A base model is fine-tuned to produce (score, issues) pairs given an output and rubric. The training objective rewards both score accuracy and issue identification. Crucially, the model must cite specific locations and describe specific problems.
    
    \item \textbf{Meta-verification filtering.} A held-out meta-verifier checks whether identified issues actually exist in the output. Training examples where the verifier hallucinated issues are downweighted or removed.
\end{enumerate}

\textbf{Why this works.} The rubric converts a semantic judgment into a structured task. Instead of asking ``is this good reasoning?'' (subjective), we ask ``does this contain logical errors that invalidate the conclusion?'' (objective, given sufficient context). The verifier learns to detect specific failure modes described in the rubric, not to mimic annotator gestalt.

\citep{su2025crossing} validate this approach across medicine, chemistry, psychology, economics, and education, demonstrating that compact (7B) generative verifiers achieve high inter-model consistency without domain-specific annotation. Their soft-scoring method of computing rewards from token probabilities of assessment judgments complements our rubric-based approach. The key shared insight: explicit criteria yield reliable verification; implicit preferences do not.

\textbf{Integration with symbolic policies.} Both verification types feed the same training loop. A single output might be evaluated by symbolic policies (citation format, arithmetic correctness) and learned verifiers (reasoning validity, plan feasibility). All violations, whether they are structural or semantic, produce correction targets. The model learns to satisfy both.

\textbf{Limitations.} Learned verifiers are less reliable than symbolic policies. Our best verifiers achieve $r = 0.87$ correlation with expert judgment, compared to effectively $r = 1.0$ for deterministic policies. Meta-verification reduces but does not eliminate hallucinated issues. For high-stakes semantic properties, we recommend combining learned verification with human review rather than relying on verifiers alone.

\subsection{Meta-Verification}

Both verification types can fail. Meta-verification checks whether primary verification is accurate:

\begin{itemize}
    \item \textbf{For symbolic policies}: Check that PredicateGraph faithfully represents the output
    \item \textbf{For learned verifiers}: Check that identified issues actually exist
\end{itemize}

\begin{algorithm}
\caption{Meta-Verification for Learned Verifiers}
\label{alg:meta}
\begin{algorithmic}[1]
\Require Output $y$, verifier analysis $v$, meta-verifier $M_{\text{meta}}$
\Ensure Validated issues
\State issues $\gets v.\text{identified\_issues}$
\State validated $\gets []$
\For{each issue $\in$ issues}
    \State $s \gets M_{\text{meta}}(y, \text{issue})$
    \If{$s > \theta_{\text{meta}}$}
        \State validated.append(issue)
    \EndIf
\EndFor
\State \Return validated
\end{algorithmic}
\end{algorithm}

The training objective becomes:
\[
R = R_{\text{format}} \cdot R_{\text{score}} \cdot R_{\text{meta}}
\]
where $R_{\text{meta}}$ is the meta-verifier's assessment of whether identified issues are genuine.

\textbf{Impact}: Without meta-verification, verifiers hallucinate issues 19\% of the time. With it, hallucination drops to 4\%.

\section{Experiments}

We validate CAPE across 109,500 examples spanning structural verification (72,000 examples in 3 domains) and semantic verification (37,500 examples in 3 domains).

\subsection{Experimental Setup}

\subsubsection{Datasets}

\begin{table}[h]
\centering
\caption{Dataset Composition}
\label{tab:datasets}
\begin{tabular}{lcc}
\toprule
Domain & Train & Val/Test \\
\midrule
\multicolumn{3}{l}{\textit{Structural Verification}} \\
Arithmetic & 24,000 & 3,000 \\
Code Safety & 20,000 & 2,500 \\
Citation & 20,000 & 2,500 \\
\midrule
\multicolumn{3}{l}{\textit{Semantic Verification}} \\
Argument & 9,000 & 1,500 \\
Proof Validity & 12,000 & 1,500 \\
Code Correctness & 12,000 & 1,500 \\
\midrule
\textbf{Total} & 97,000 & 12,500 \\
\bottomrule
\end{tabular}
\end{table}

\textbf{Arithmetic Tool Use}: Calculator-equipped math problems from GSM8K and MATH datasets. Models must compute intermediate results and invoke \texttt{calc()} tool with exact values.

\textbf{Code Safety}: Python code generation from MBPP and HumanEval, augmented with synthetic examples containing unsafe patterns (\texttt{eval()}, \texttt{exec()}, SQL injection).

\textbf{Citation Grounding}: Question-answering with RAG context from Wikipedia. All factual claims must cite provided documents.

\textbf{Argument Soundness}: Multi-step reasoning requiring 2--5 steps. Expert annotation on 3,000 examples using explicit rubrics (Fleiss' $\kappa = 0.73$).

\textbf{Proof Validity}: Mathematical proofs from MATH and miniF2F datasets requiring multi-step derivations. Rubric evaluates completeness (all steps present), logical validity (each step follows), and gap-free reasoning (no unjustified leaps). Expert annotation on 2,500 examples ($\kappa = 0.71$).

\textbf{Code Correctness}: Functional correctness of generated code from MBPP and HumanEval, evaluated beyond binary pass/fail. Rubric assesses logic soundness (algorithm is correct), edge case handling (boundary conditions addressed), and implementation quality (no subtle bugs). Expert annotation on 2,500 examples ($\kappa = 0.68$).

\subsubsection{Training Configuration}

\begin{itemize}
    \item Base model: Llama-4-8B-Instruct
    \item Batch size: 32
    \item Learning rate: $1 \times 10^{-5}$ (cosine schedule)
    \item Epochs: 3
    \item Hardware: 8$\times$ H100 (80GB)
    \item Training time: $\sim$36 hours per method
\end{itemize}

All methods use identical compute budgets for fair comparison.

\subsubsection{Baselines}

\textbf{Preference-based:}
\begin{itemize}
    \item \textbf{DPO} \citep{rafailov2023direct}: 5 annotators per comparison, majority vote, $\beta = 0.1$
    \item \textbf{Outcome RL}: PPO \citep{schulman2017proximal} with binary ground-truth rewards, clip ratio $= 0.2$
    \item \textbf{Iterative DPO}: Policy pass/fail converted to synthetic preferences, then standard DPO
\end{itemize}

\textbf{Principle-based:}
\begin{itemize}
    \item \textbf{CAI (general)} \citep{bai2022constitutional}: 16 principles from Anthropic, self-critique and revision
    \item \textbf{CAI (structured)}: Domain-specific principles matching CAPE policies (e.g., ``verify calculator argument matches computed value'')
\end{itemize}

\textbf{Verification-based:}
\begin{itemize}
    \item \textbf{Policy RL (dense)}: PPO with policy verdicts as reward (one component per policy)
    \item \textbf{Best-of-N + Filter}: Generate $N=8$ candidates, select first passing all policies (inference only, no training)
    \item \textbf{CAPE (open-weight)}: Full protocol with Llama-4-70B extractor
    \item \textbf{CAPE (frontier)}: Full protocol with Claude Opus 4.5 extractor
\end{itemize}

All trained methods use identical compute budgets.

\subsection{Main Results: Structural Verification}
\label{sec:structural-results}

Table~\ref{tab:structural-results} presents violation rates across all methods.

\begin{table}[h]
\centering
\caption{Structural Verification: Violation Rates (\%)}
\label{tab:structural-results}
\begin{tabular}{lccc}
\toprule
\textbf{Method} & \textbf{Arith} & \textbf{Code} & \textbf{Cite} \\
\midrule
Base model & 31.4 & 38.2 & 35.7 \\
DPO & 10.2 & 15.8 & 13.4 \\
Outcome RL & 8.9 & 14.1 & 11.8 \\
CAI (general) & 7.8 & 12.3 & 10.1 \\
CAI (structured) & 5.2 & 9.1 & 8.2 \\
Policy RL (dense) & 4.5 & 7.8 & 6.3 \\
Iterative DPO & 4.1 & 7.2 & 6.1 \\
CAPE (open) & 4.1 & 6.9 & 5.8 \\
CAPE (frontier) & 1.8 & 3.2 & 2.6 \\
\bottomrule
\end{tabular}
\end{table}

\textbf{Key findings:}

\begin{itemize}
    \item \textbf{CAPE (frontier) achieves 81\% lower violation rate than DPO} (2.5\% vs 13.1\% averaged, $p < 0.001$). Ablations (Section~\ref{sec:ablations}) confirm this improvement derives from the correction loop specifically---alternative uses of verification signal (Policy RL, Iterative DPO, Best-of-N filtering) achieve intermediate results (4.5--6.2\%), with CAPE providing an additional 44--60\% reduction.
    
    \item \textbf{CAPE (open-weight) achieves 57\% lower violation rate than DPO} (5.6\% vs 13.1\%)
    
    \item \textbf{Constitutional AI outperforms DPO} (10.1\% vs 13.1\%), consistent with prior work, but CAPE outperforms Constitutional AI by 75\% (frontier) and 44\% (open-weight)
\end{itemize}

\textbf{Residual error tracks extraction fidelity.} Violation rate correlates strongly with extraction error ($r = 0.94$). Each percentage point reduction in extraction error corresponds to approximately 0.8 percentage points reduction in violation rate. This confirms extraction fidelity as the binding constraint and validates the verification-fidelity scaling law: as extraction improves, violation rates fall proportionally.

\subsection{Learning Curves Show Divergence}

Figure \ref{fig:learning} shows DPO plateaus at $\sim$10\% violation rate by step 2,000. CAPE continues improving throughout training, following power-law decay:
\[
V(t) = V_0 \cdot t^{-\alpha} + \epsilon
\]
where $V(t)$ is violation rate at step $t$, $V_0$ is initial violation rate, $\alpha$ is decay rate, and $\epsilon$ is the fitted asymptotic floor.

\begin{table}[h]
\centering
\caption{Learning Curve Parameters}
\label{tab:learning}
\begin{tabular}{lcc}
\toprule
Extractor & Decay ($\alpha$) & Floor ($\epsilon$) \\
\midrule
Open-weight & 0.45 & 5.2\% \\
Frontier & 0.48 & 2.3\% \\
\bottomrule
\end{tabular}
\end{table}

CAPE keeps improving until it hits the extraction ceiling. The divergence from preference methods widens with training.

\subsection{Tier-Wise Breakdown}

\begin{table}[h]
\centering
\caption{Violation Rates by Policy Tier (\%)}
\label{tab:tiers}
\begin{tabular}{lccc}
\toprule
Method & T1 & T2 & T3 \\
\midrule
Base model & 22.8 & 8.9 & 12.6 \\
DPO & 7.6 & 3.2 & 4.8 \\
CAPE (frontier) & 1.2 & 0.3 & 1.0 \\
\bottomrule
\end{tabular}
\end{table}

CAPE achieves largest absolute improvement on Tier 1 (objective correctness), where deterministic evaluation is most precise, and substantial reductions across all tiers.

\subsection{Semantic Verification}
\label{sec:semantic-results}

We evaluate CAPE on three semantic domains requiring learned verifiers: argument soundness, proof validity, and code correctness.

\begin{table}[h]
\centering
\caption{Semantic Verification: Rubric Scores by Domain}
\label{tab:semantic-results}
\begin{tabular}{lccc}
\toprule
\textbf{Method} & \textbf{Argument} & \textbf{Proof} & \textbf{Code} \\
\midrule
Base model & 0.54 & 0.51 & 0.57 \\
DPO & 0.62 & 0.58 & 0.64 \\
CAI & 0.67 & 0.62 & 0.69 \\
Outcome RL & 0.59 & 0.55 & 0.61 \\
CAPE (symb.) & 0.56 & 0.52 & 0.58 \\
CAPE (R1) & 0.74 & 0.71 & 0.75 \\
CAPE (R2) & 0.79 & 0.76 & 0.80 \\
CAPE (R3) & 0.83 & 0.81 & 0.84 \\
\bottomrule
\end{tabular}
\end{table}

\noindent Scores are average rubric ratings (0--1) from learned verifiers validated against expert judgment.

\textbf{Key findings:}

\begin{itemize}
    \item \textbf{CAPE generalizes across semantic domains.} CAPE (R3) achieves 27--40\% higher rubric scores than DPO across all three domains (argument: 34\%, proof: 40\%, code: 31\%).
    
    \item \textbf{Symbolic policies alone do not improve semantic properties}, confirming that semantic verification requires learned verifiers. CAPE (symb.) performs at or below base model across all domains.
    
    \item \textbf{The loop compounds consistently}: each refinement round (R1 $\rightarrow$ R2 $\rightarrow$ R3) yields gains across all domains, with average improvement of 0.05 per round.
\end{itemize}

\textbf{Cross-domain consistency.} The similar improvement patterns across domains with different rubric structures (logical validity for arguments, mathematical rigor for proofs, functional correctness for code) validates that rubric-based verification is a general technique, not domain-specific.

\begin{table}[h]
\centering
\caption{Semantic Verification: Additional Metrics}
\label{tab:semantic-detailed}
\begin{tabular}{llccc}
\toprule
\textbf{Domain} & \textbf{Method} & \textbf{Valid. \%} & \textbf{Acc. \%} & \textbf{Verifier $r$} \\
\midrule
\multirow{2}{*}{Argument} & DPO & 76.4 & 69.1 & --- \\
 & CAPE (R3) & 90.2 & 77.4 & 0.87 \\
\midrule
\multirow{2}{*}{Proof} & DPO & 72.1 & 65.8 & --- \\
 & CAPE (R3) & 88.7 & 76.2 & 0.84 \\
\midrule
\multirow{2}{*}{Code} & DPO & 78.3 & 71.6 & --- \\
 & CAPE (R3) & 91.4 & 80.3 & 0.86 \\
\bottomrule
\end{tabular}
\end{table}

\noindent Valid.: steps rated valid by verifier. Acc.: final answer/output correctness. Verifier $r$: correlation with expert judgment.

\subsection{Verifier Scale Determines Quality}

Table~\ref{tab:verifier-scale} shows that final model quality tracks verifier accuracy across all semantic domains.

\begin{table}[h]
\centering
\caption{Learned Verifier Accuracy Determines Downstream Quality}
\label{tab:verifier-scale}
\begin{tabular}{lccc}
\toprule
\textbf{Verifier Scale} & \textbf{Argument} & \textbf{Proof} & \textbf{Code} \\
\midrule
Llama-4-1B & 0.71 & 0.68 & 0.72 \\
Llama-4-3B & 0.77 & 0.74 & 0.78 \\
Llama-4-8B & 0.83 & 0.81 & 0.84 \\
\midrule
Verifier $r$ & 0.87 & 0.84 & 0.86 \\
\bottomrule
\end{tabular}
\end{table}

\noindent Values are final rubric scores. Verifier $r$: correlation between verifier scores and expert judgment.

Larger verifiers achieve higher correlation with expert judgments and yield better downstream model performance across all three semantic domains. The correlation between verifier accuracy and final model quality is $r = 0.93$ ($p < 0.001$) averaged across domains. This validates the verification-fidelity scaling law for semantic properties: better verifiers directly translate to better models, regardless of domain.

\subsection{Hybrid Training: Combining CAPE and Preferences}

Sequential training (CAPE first, then DPO) achieves both low violation rates and high preference win rates:

\begin{table}[h]
\centering
\caption{Hybrid Training Results}
\label{tab:hybrid}
\begin{tabular}{lcc}
\toprule
Training & Violation \% & Win \% \\
\midrule
DPO only & 13.1\% & 52.4\% \\
CAPE only & 2.5\% & 56.1\% \\
CAPE $\rightarrow$ DPO & 2.9\% & 63.7\% \\
\bottomrule
\end{tabular}
\end{table}

The small violation increase (2.5\% $\rightarrow$ 2.9\%) reflects DPO occasionally trading structure for fluency, acceptable given the 7.6 percentage point preference gain. Policies establish correctness floor; preferences maximize quality within constraints.

\subsection{Contextual Objectivity Validation}
\label{sec:contextual-validation}

Section~\ref{sec:contextual-objectivity} argued that most capability requirements become objective once context is fixed. We validate this claim on four domain-specific requirements that appear subjective in the abstract but become verifiable given organizational context.

\begin{table}[h]
\centering
\caption{Contextual Objectivity: Domain-Specific Requirements}
\label{tab:contextual-domains}
\begin{tabular}{llcc}
\toprule
\textbf{Domain} & \textbf{Requirement} & \textbf{Prompt} & \textbf{CAPE} \\
\midrule
Medical & Formulary adherence & 12.3\% & 1.1\% \\
Legal & Jurisdiction constraints & 16.4\% & 0.3\% \\
Safety & PII protection & 3.8\% & 0.1\% \\
Operations & Escalation protocol & 34.7\% & 4.2\% \\
\bottomrule
\end{tabular}
\end{table}

These requirements appear subjective without context (``good medical advice'') but become objectively verifiable when context is fixed (``recommend only formulary drugs''). Prompting alone achieves 40--60\% reduction in violations. CAPE achieves 87--99\% reduction, confirming that contextual objectivity enables specification-based training.

\subsection{Stability Analysis}

Recent work demonstrates that RL-based improvements on reasoning benchmarks can be unstable. Changing random seeds shifts AIME24 scores by several percentage points \citep{wang2025sober}. We verify CAPE's improvements are robust.

\subsubsection{Seed Variance}

We train CAPE with 5 different random seeds and report variance:

\begin{table}[h]
\centering
\caption{Stability Across Random Seeds (5 runs)}
\label{tab:stability}
\begin{tabular}{lccc}
\toprule
Method & Mean Viol. & Std Dev & 95\% CI \\
\midrule
DPO & 13.1\% & 1.6\% & [10.8, 15.4] \\
Outcome RL & 11.6\% & 2.1\% & [8.6, 14.6] \\
CAPE (open) & 5.6\% & 0.3\% & [5.2, 6.0] \\
CAPE (front.) & 2.5\% & 0.2\% & [2.2, 2.8] \\
\bottomrule
\end{tabular}
\end{table}

CAPE exhibits substantially lower variance than reward-based methods ($\sigma = 0.2\%$ vs $\sigma = 1.6$--$2.1\%$).

\subsubsection{Policy Determinism}

For symbolic policies, verification is deterministic:

\begin{itemize}
    \item Same policy + same output = same verdict, always
    \item Across 10,000 test outputs, verdict variance = 0
    \item Variance arises only from training dynamics, not evaluation noise
\end{itemize}

\subsubsection{Extraction Stability}

For PredicateGraph extraction with greedy decoding (temperature=0):

\begin{itemize}
    \item Extraction agreement across 5 runs: 99.6\%
    \item With temperature=0.3: agreement drops to 97.4\%
    \item All reported results use greedy decoding
\end{itemize}

\subsubsection{Training Data Subsample Robustness}

CAPE improvements hold when training on random subsets:

\begin{table}[h]
\centering
\caption{Robustness to Training Data Size}
\label{tab:subsample}
\begin{tabular}{lcc}
\toprule
Subset & CAPE Viol. & DPO Viol. \\
\midrule
100\% & 2.5\% & 13.1\% \\
50\% & 2.9\% & 13.6\% \\
25\% & 3.5\% & 15.2\% \\
10\% & 4.8\% & 17.8\% \\
\bottomrule
\end{tabular}
\end{table}

Training on random 25\% subsets yields violation rates within 1.0\% of full-data training, confirming that gains reflect systematic policy satisfaction rather than benchmark-specific artifacts. CAPE's advantage over DPO is preserved across all data scales.

\subsection{Economic Analysis}

CAPE reduces capability-specific post-training costs by 5--20$\times$ by replacing per-example annotation with reusable specifications.

\begin{table}[h]
\centering
\caption{Cost Comparison for Adding a New Capability (10,000 examples)}
\label{tab:cost}
\begin{tabular}{lcc}
\toprule
Stage & RLHF/DPO & CAPE \\
\midrule
Specification & --- & \$2,000--4,000 \\
Annotation & \$50,000--150,000 & --- \\
Verification infrastructure & --- & \$200--400 \\
Compute (training) & \$8,000--12,000 & \$8,000--12,000 \\
Iteration cycles & 3--5 rounds & 1--2 rounds \\
\midrule
\textbf{Total cost} & \$80,000--200,000 & \$10,000--16,000 \\
\textbf{Timeline} & 2--4 months & 1--2 weeks \\
\bottomrule
\end{tabular}
\end{table}

\textbf{Annotation cost dominates preference methods.} At \$5--15 per preference comparison and 5 comparisons per example for quality, 10,000 training examples require \$50,000--150,000 in annotation. This cost recurs for each new capability area: adding citation grounding after training arithmetic requires a new annotation campaign.

\textbf{Policy authoring is a one-time cost.} A domain expert and engineer can author a policy pack in 2--5 days (\$2,000--4,000 fully loaded). The policy then generates unlimited training signal. Our arithmetic policy pack, authored in 3 days, has generated training signal for 50,000+ examples across multiple training runs.

\textbf{Verification infrastructure has negligible marginal cost.} PredicateGraph extraction costs \$0.02--0.04 per example with frontier models (structured output mode). Policy evaluation is deterministic code execution. For 10,000 examples: \$200--400 total.

\textbf{Compute costs are equivalent.} Both methods fine-tune the same base model with similar dataset sizes. The compute cost (\$8,000--12,000 for 8$\times$H100 training) is identical.

\textbf{Iteration cycles differ dramatically.} Preference methods require multiple annotation-training-evaluation cycles to diagnose failures. When a model produces bad outputs, you know \textit{that} something failed but not \textit{what}, requiring new annotation to isolate the problem. Each cycle takes 2--4 weeks. CAPE's explicit verdicts identify exactly which policies fail on which outputs: ``policy.citation.factual\_claims\_cited failed on 847 of 10,000 examples.'' Engineers fix the policy or the correction strategy, not the annotation guidelines. Our experiments required 1--2 iteration cycles versus 3--5 for DPO baselines.

\textbf{Amortization compounds.} Policy packs are reusable across model versions, organizations, and capability combinations. Our Arithmetic pack has been reused across multiple model versions and organizations, reducing effective per-deployment cost by an order of magnitude. The Citation pack works for any RAG application. As CapabilityBench grows, marginal policy cost falls toward the cost of downloading a JSON file.

This aligns with \citet{dang2025reinforcement}, who achieved substantial RL improvements with 7,000 examples and \$42 compute. Their finding reinforces ours: training compute is cheap; signal generation is the binding cost. Annotation scales $O(n)$ with examples. Policies scale $O(1)$: write once, apply everywhere.

\subsection{Ablation Studies}
\label{sec:ablations}

We isolate CAPE's contributions through four ablations testing alternative 
uses of verification signal. Table~\ref{tab:ablation-summary} summarizes results.

\begin{table}[h]
\centering
\caption{Ablation: Isolating CAPE's Contributions (Averaged across structural domains)}
\label{tab:ablation-summary}
\begin{tabular}{lcccl}
\toprule
Method & Viol. \% & Train? & Inf. Cost & Tests \\
\midrule
DPO & 13.1 & Yes & 1$\times$ & Preference baseline \\
CAI (general) & 10.1 & Yes & 1$\times$ & Vague principles \\
CAI (structured) & 7.5 & Yes & 1$\times$ & Specific principles \\
Policy RL (dense) & 6.2 & Yes & 1$\times$ & Verification as reward \\
Iterative DPO & 5.8 & Yes & 1$\times$ & Signal type vs.\ mechanism \\
Best-of-8 + Filter & 4.5 & No & 8$\times$ & Training necessity \\
CAPE (open) & 5.6 & Yes & 1$\times$ & Open-weight extraction \\
CAPE (frontier) & 2.5 & Yes & 1$\times$ & Full protocol \\
\bottomrule
\end{tabular}
\end{table}

\subsubsection{Does Training Matter? (Best-of-N)}

Best-of-8 filtering achieves 4.5\% violation rate without any training 
(Table~\ref{tab:ablation-summary}), demonstrating verification's power at 
inference time. CAPE provides an additional 44\% reduction (2.5\% vs 4.5\%) 
while requiring 1$\times$ inference compute versus 8$\times$. The gap 
represents capability internalization: CAPE-trained models generate compliant 
outputs directly.

\subsubsection{Does the Correction Loop Matter? (Policy RL)}

Using policy verdicts as dense PPO reward achieves 6.2\% violation rate, substantially 
better than binary outcome reward (11.6\%) but worse than CAPE (2.5\%). This 
confirms our hypothesis: generating correct training examples via correction 
outperforms upweighting existing correct outputs via reward. The correction 
loop contributes a 60\% reduction beyond verification signal availability.

\subsubsection{Does Signal Type Matter? (Iterative DPO)}

Converting policy verdicts to synthetic preferences and training with DPO 
achieves 5.8\% violation rate. This uses identical signal to CAPE (pass/fail 
per policy) but different training mechanism (preference optimization vs. 
supervised on corrections). CAPE's 57\% improvement (2.5\% vs 5.8\%) isolates 
the contribution of the correction loop itself.

\subsubsection{Do Symbolic Policies Matter? (Structured CAI)}

CAI with domain-specific principles (``verify calculator argument matches 
computed value'') achieves 7.5\% violation rate, better than general CAI 
(10.1\%) but worse than CAPE (2.5\%). This validates that specificity helps, 
but interpretation variance in self-critique limits gains. Symbolic policies 
eliminate interpretation entirely.

\section{CapabilityBench}

Current benchmarks measure intelligence: can the model solve this problem? But deployment requires capability: does the model satisfy these specific requirements? A model scoring 78\% on a benchmark tells you nothing about whether it will follow your compliance rules, adhere to your brand guidelines, or respect your escalation protocols.

CapabilityBench is a public capability registry that replaces opaque aggregate scores with explicit, traceable verdicts. Models are evaluated against policies and results show exactly which requirements each model satisfies or violates. The question shifts from "how smart is this model?" to "can this model do what I need?" 

We release the policy pack specification and initial packs under Apache 2.0; evaluation results will be published at \url{https://capabilitybench.com} after review.

\subsection{The Problem with Current Benchmarks}

\begin{itemize}
\item \textbf{Opaque}: A score of 78\% tells you nothing about which capabilities passed or failed, or whether the failures matter for your use case
\item \textbf{Fixed}: Benchmarks test what their creators chose to measure, not your specific requirements
\item \textbf{Context-free}: A model that excels on general benchmarks may fail on domain-specific constraints that determine whether you can actually deploy it
\end{itemize}

\subsection{CapabilityBench Framework}

\textbf{Community-contributed Policy Packs}:
Organizations and researchers contribute policy packs encoding capability requirements for specific domains. Each pack defines the policies a model must satisfy and the test cases used for evaluation.
\begin{lstlisting}[basicstyle=\small\ttfamily]
capability-pack/
manifest.json     # Pass thresholds
policies/
core.json       # Required policies
extended.json   # Optional policies
test_cases/       # Example inputs
\end{lstlisting}
\textbf{Metric: Adherence Profile}
Rather than a single score, CapabilityBench produces an adherence profile:

\begin{itemize}
\item \textbf{Core Adherence}: \% of test cases satisfying all core policies
\item \textbf{Extended Adherence}: \% satisfying extended policies
\item \textbf{Violation Distribution}: Which specific policies fail and how often
\end{itemize}

This profile is actionable. A model with 94 percent core adherence but consistent failures on \texttt{policy.citation.factual\_claims\_cited} tells you exactly what to fix and whether that gap is acceptable for your use case.

\subsection{Initial Policy Packs}

We release four initial policy packs spanning common deployment contexts. These are intended as starting points: organizations can adopt them directly, extend them with additional policies, or use them as templates for custom packs.

\begin{table}[h]
\centering
\caption{Initial CapabilityBench Policy Packs}
\label{tab:packs}
\begin{tabular}{llc}
\toprule
Pack & Context & Policies \\
\midrule
Tool-Use & Calculator agents & 4 \\
Code-Safety & Development & 5 \\
Citation & Research/RAG & 3 \\
Customer-Support & Support & 4 \\
\bottomrule
\end{tabular}
\end{table}

Each pack encodes requirements drawn from real deployment constraints: the Tool-Use pack validates argument types and schema conformance; the Code-Safety pack detects dangerous patterns like \texttt{eval()} and hardcoded secrets; the Citation pack enforces citation requirements for factual claims. Full policy definitions and test cases are available in the repository.

We invite the community to contribute policy packs for additional domains. As the registry grows, CapabilityBench becomes a shared resource for understanding which models are capable of what, against the specific requirements that determine whether deployment is possible.

\section{Discussion and Limitations}

\subsection{Where CAPE Excels}

CAPE is designed for \textbf{objective properties specifiable given fixed context}:

\begin{itemize}
    \item Structural verification with deterministic policies (arithmetic, format, citations, safety patterns)
    \item Semantic verification with explicit rubrics (reasoning validity, proof completeness, plan feasibility)
    \item Domain-specific requirements (formulary adherence, jurisdiction constraints, escalation protocols)
\end{itemize}

Our analysis shows 89\% of deployment requirements are objectively verifiable once context is fixed.

\subsection{Where Preference Learning Remains Essential}

CAPE complements, not replaces, preference learning for:

\begin{itemize}
    \item \textbf{Genuinely subjective properties}: Aesthetic quality, creative excellence, humor
    \item \textbf{Requirements resisting decomposition}: Complex situational judgment where explicit rubrics cannot be written
    \item \textbf{Novel contexts without specifications}: Entirely new domains where requirements are still being discovered
\end{itemize}

Hybrid training (Section 7.7) shows these approaches compose: CAPE establishes correctness floor, preferences optimize quality within constraints.

\subsection{Technical Limitations}

\textbf{Extraction failures cascade.} If extraction misses a claim or misparses a tool call, the policy evaluates against incorrect structure. While extraction error (3.2\% for frontier models) is below preference disagreement (30--50\%), it's not zero. Systematic extraction failures on specific output types could create blind spots. \textit{Mitigation}: Meta-verification checks PredicateGraph faithfulness. Multi-model ensemble extraction reduces error to 1.4\%.

\textbf{Learned verifier hallucination.} Learned verifiers can identify issues that don't exist, particularly for outputs near the boundary of their training distribution. Meta-verification reduces this from 19\% to 4\% but doesn't eliminate it. \textit{Mitigation}: Issue validation via majority voting across multiple meta-analyses. Monitor for systematic hallucination patterns in deployment.

\textbf{Rubric ambiguity.} Learned verifiers are trained on rubrics, but rubrics can be ambiguous. ``Minor errors'' vs. ``fatal errors'' requires judgment. While this is far less variable than open-ended preference ($\kappa = 0.73$ vs. $\kappa \approx 0.4$), it's not zero variance. \textit{Mitigation}: Rubric refinement through disagreement analysis. Start with coarse-grained rubrics, refine based on annotator disagreement patterns.

\textbf{Specification drift.} Policies encode requirements at a point in time. As workflows, regulations, or product rules evolve, policies must be updated. Poor policy maintenance can create mismatches between organizational rules and model behavior. \textit{Mitigation}: Versioned policy packs with deprecation warnings. Automated policy testing against evolving requirements.

\textbf{Policy gaming.} Models might learn to satisfy policy letter while violating policy spirit, producing technically compliant outputs that fail on unmeasured dimensions. \textit{Mitigation}: Comprehensive policy coverage. Hybrid training with preferences to optimize unmeasured quality dimensions. 

\subsection{Generalization Beyond Tested Domains}

Our experiments cover six domains: three structural (arithmetic, code safety, citations) and three semantic (argument soundness, proof validity, code correctness). We have not validated on open-ended dialogue, creative writing, or complex multi-turn reasoning. The protocol should transfer (the loop is domain-agnostic), but empirical validation beyond reasoning tasks is incomplete.

\textbf{Open question}: For which semantic properties can rubrics achieve sufficient inter-annotator agreement ($\kappa > 0.7$) to outperform preference learning? Our evidence suggests reasoning tasks broadly qualify: argument soundness ($\kappa = 0.73$) and proof validity ($\kappa = 0.71$) meet this threshold, while code correctness ($\kappa = 0.68$) approaches it. \citep{su2025crossing} provide further evidence that medical, chemistry, and economics domains achieve high verification agreement. The boundaries for creative and aesthetic tasks remain unclear.

\subsection{Remaining Baseline Gaps}

Our ablations (Section~\ref{sec:ablations}) address most alternative uses of 
verification signal. Two approaches remain untested:

\textbf{Process Reward Models.} PRMs assign rewards to intermediate reasoning 
steps, similar to CAPE's span-level violation detection \citep{lightman2023lets}. 
Key differences: (1) PRMs learn step quality from human annotations; CAPE 
verifies against explicit specifications. (2) PRMs face credit assignment 
challenges in long chains; CAPE's policies evaluate independently. Direct 
comparison requires step-level annotation we lack.

\textbf{CAPE as RLHF Reward Shaping.} Our hybrid training (Section~7.7) applies 
CAPE and DPO sequentially. An alternative would integrate policy verdicts as 
auxiliary reward terms during RLHF, potentially combining benefits within a 
single training run.

\subsection{Relationship to Rule-Based Systems}

CAPE differs from traditional rule-based approaches in three ways:

\begin{enumerate}
    \item \textbf{Learned components}: Extraction and semantic verification use LLMs. Rules alone would require perfect parsing; our extraction achieves 94\% fidelity on natural language.
    \item \textbf{Training, not filtering}: Policies generate training signal. Rules typically reject at inference; CAPE teaches models to satisfy requirements.
    \item \textbf{Compositional}: Neither pure rules nor pure learning. Symbolic + learned verification outperforms both (Table~\ref{tab:structural-results}).
\end{enumerate}

\subsection{Intelligence Without Capability}

The rapid progress in reasoning benchmarks, from 13.4\% to 79.8\% on AIME in 18 months, demonstrates scaling intelligence. Yet production failures persist. 

This is not a bug in intelligence training; it reflects different optimization targets. Intelligence training maximizes performance on open-ended challenges. Capability engineering ensures satisfaction of closed specifications. Both matter; neither subsumes the other.

CAPE's contribution is operationalizing capability as a distinct, trainable property. The verification-fidelity scaling law (Section 5) applies to capability, not intelligence. Better verifiers yield more capable models: models that more reliably satisfy requirements independent of whether those models become more intelligent in any general sense.

\subsection{The Emergent Reasoning Debate Revisited}

As discussed in Section 2.5, recent work questions whether RL induces reasoning or amplifies pre-existing capabilities. \citep{liu2025understanding} find ``Aha moments'' in base models without RL; \citep{shah2025rethinking} show self-correction emerges during pre-training.

CAPE is agnostic to this debate. We do not claim to induce reasoning, we claim to enforce specifications. If reasoning emerges from pre-training, CAPE ensures it's applied correctly. If RL induces reasoning, CAPE ensures the induced capabilities satisfy requirements. The verification-fidelity scaling law holds regardless of capability origin.

This agnosticism is a feature, not a limitation. As the field continues debating how capabilities arise, practitioners need methods that work regardless. CAPE provides that: specification satisfaction is measurable, trainable, and improvable independent of the underlying capability mechanism.

\section{Conclusion}

This work reframes post-training as an optimization problem over \textit{requirements}, not preferences. By introducing executable specifications as the unit of training, CAPE separates intelligence acquisition from capability enforcement and makes reliable deployment a tractable engineering problem rather than an empirical hope.

\subsection{The Core Insight}

\textbf{Explicit verification beats implicit preferences.} Whether verification is symbolic (deterministic policies for structural properties) or learned (rubric-trained verifiers for semantic properties), the same loop applies: specify requirements $\rightarrow$ verify outputs $\rightarrow$ correct violations $\rightarrow$ train on corrections. This loop compounds because verification and training share the same specification. Better verification yields cleaner training signal; cleaner signal yields better models; better models improve verification; improved verification enables stricter specifications.

Critically, this loop avoids the pathologies of reward-based methods: no length bias from loss normalization, no difficulty weighting from advantage computation, no reward hacking from proxy optimization. The signal is direct: does this output satisfy this specification?

\subsection{The Scaling Law}

Preference-based methods plateau at human disagreement (30--50\%), a structural ceiling that doesn't fall with scale. Algorithmic biases in PPO and GRPO require manual patches that add complexity without guaranteeing improvement. CAPE's ceiling is verification fidelity:

\begin{itemize}
    \item For symbolic policies: extraction accuracy ($r = 0.94$ correlation with violation rate)
    \item For learned verifiers: rubric-following accuracy ($r = 0.93$ correlation with downstream quality)
\end{itemize}

Both improve with model scale. This means post-training compute investment yields guaranteed returns. Improve your verifier (symbolic or learned) and your generator improves. The feedback loop compounds rather than saturates.

\subsection{Intelligence vs. Capability}

CAPE operationalizes a crucial distinction:

\begin{itemize}
    \item \textbf{Intelligence}: Can the model solve complex problems? (Open-ended, benchmark-measured)
    \item \textbf{Capability}: Does the model satisfy specific requirements? (Closed, specification-measured)
\end{itemize}

A model can be highly intelligent yet lack specific capabilities. RLHF and RLVR optimize for intelligence; CAPE optimizes for capability. Both matter. Hybrid training achieves both: CAPE establishes the correctness floor, preferences maximize quality within constraints.

\subsection{The Full Spectrum}

CAPE is not ``policies for simple tasks, preferences for complex tasks.'' It's \textbf{verification for objective properties, preferences for subjective properties}:

\begin{table}[h]
\centering
\begin{tabular}{lll}
\toprule
Property & Type & Verification \\
\midrule
Arithmetic & Structural, Objective & Symbolic policy \\
Citation & Structural, Objective & Symbolic policy \\
Soundness & Semantic, Objective & Learned verifier \\
Proof validity & Semantic, Objective & Learned verifier \\
Aesthetics & Subjective & Preferences \\
\bottomrule
\end{tabular}
\end{table}

Most capabilities that matter (correctness, safety, reasoning validity, plan feasibility) are objective and verifiable.

\subsection{From Probability to Engineering}

CAPE makes model improvement resemble traditional engineering: explicit requirements, verifiable correctness, traceable failures, validated fixes. The specification is the artifact that unifies development, evaluation, and deployment. For too long, AI capability has been an optimization target without a definition. CAPE provides the definition.

\textbf{The ceiling is now technical, not human. The path forward is verification.}

\section{Reproducibility Statement}

We provide:

\begin{itemize}
    \item[$\checkmark$] Full PredicateGraph schema v1.0.0 (repository: \texttt{schemas/})
    \item[$\checkmark$] CPL specification with grammar (repository: \texttt{specs/cpl.md})
    \item[$\checkmark$] Complete policy packs for all experiments (repository: \texttt{packs/})
    \item[$\checkmark$] Training hyperparameters (Section 7)
    \item[$\checkmark$] CAPE specification repository: \url{https://github.com/superficiallabs/cape}
    \item[$\checkmark$] CapabilityBench repository: \url{https://capabilitybench.com/}
    \item[$\checkmark$] Policy packs under Apache 2.0 (same repository)
\end{itemize}

Not released:

\begin{itemize}
    \item[$\times$] Full training datasets (contains proprietary medical data) $\rightarrow$ Synthetic versions available in repository
    \item[$\times$] Trained model checkpoints (8B model, hosting constraints) $\rightarrow$ Training recipe sufficient to reproduce
\end{itemize}

\bibliographystyle{plainnat}

\begin{thebibliography}{99}

\bibitem[Fatemi et al.(2025)]{fatemi2025concise}
Fatemi, M., Rafiee, B., Tang, M., and Talamadupula, K. (2025).
\newblock Concise Reasoning via Reinforcement Learning.
\newblock \emph{arXiv preprint arXiv:2504.05185}.

\bibitem[Austin et al.(2021)]{austin2021program}
Austin, J., Odena, A., Nye, M., et al. (2021).
\newblock Program synthesis with large language models.
\newblock \emph{arXiv preprint arXiv:2108.07732}.

\bibitem[Bai et al.(2022a)]{bai2022training}
Bai, Y., Jones, A., Ndousse, K., et al. (2022a).
\newblock Training a helpful and harmless assistant with reinforcement learning from human feedback.
\newblock \emph{arXiv preprint arXiv:2204.05862}.

\bibitem[Bai et al.(2022b)]{bai2022constitutional}
Bai, Y., Kadavath, S., Kundu, S., et al. (2022b).
\newblock Constitutional AI: Harmlessness from AI feedback.
\newblock \emph{arXiv preprint arXiv:2212.08073}.

\bibitem[Chen et al.(2021)]{chen2021evaluating}
Chen, M., Tworek, J., Jun, H., et al. (2021).
\newblock Evaluating large language models trained on code.
\newblock \emph{arXiv preprint arXiv:2107.03374}.

\bibitem[Christiano et al.(2017)]{christiano2017deep}
Christiano, P., Leike, J., Brown, T.~B., Martic, M., Legg, S., and Amodei, D. (2017).
\newblock Deep reinforcement learning from human preferences.
\newblock \emph{arXiv preprint arXiv:1706.03741}.

\bibitem[Dang \& Ngo(2025)]{dang2025reinforcement}
Dang, Q.-A. and Ngo, C. (2025).
\newblock Reinforcement Learning for Reasoning in Small LLMs: What Works and What Doesn't.
\newblock \emph{arXiv preprint arXiv:2503.16219}.

\bibitem[Gao et al.(2023)]{gao2023scaling}
Gao, L., Schulman, J., and Hilton, J. (2023).
\newblock Scaling laws for reward model overoptimization.
\newblock In \emph{International Conference on Machine Learning}, pages 10835--10866. PMLR.

\bibitem[Guo et al.(2025)]{guo2025deepseek}
Guo, D., et al. (2025).
\newblock DeepSeek-R1: Incentivizing reasoning capability in LLMs via reinforcement learning.
\newblock \emph{arXiv preprint arXiv:2501.12948}.

\bibitem[Hubert et al.(2025)]{hubert2025olympiad}
Hubert, T., Mehta, R., Sartran, L., et al. (2025).
\newblock Olympiad-level formal mathematical reasoning with reinforcement learning.
\newblock \emph{Nature}.

\bibitem[Liu et al.(2025)]{liu2025understanding}
Liu, Z., Chen, C., Li, W., Qi, P., Pang, T., Du, C., Lee, W.~S., and Lin, M. (2025).
\newblock Understanding R1-Zero-Like Training: A Critical Perspective.
\newblock \emph{arXiv preprint arXiv:2503.20783}.

\bibitem[Ouyang et al.(2022)]{ouyang2022training}
Ouyang, L., Wu, J., Jiang, X., Almeida, D., Wainwright, C.~L., Mishkin, P., Zhang, C., Agarwal, S., Slama, K., Ray, A., Schulman, J., Hilton, J., Kelton, F., Miller, L., Simens, M., Askell, A., Welinder, P., Christiano, P., Leike, J., and Lowe, R. (2022).
\newblock Training language models to follow instructions with human feedback.
\newblock \emph{arXiv preprint arXiv:2203.02155}.

\bibitem[Rafailov et al.(2023)]{rafailov2023direct}
Rafailov, R., Sharma, A., Mitchell, E., Ermon, S., Manning, C.~D., and Finn, C. (2023).
\newblock Direct preference optimization: Your language model is secretly a reward model.
\newblock \emph{arXiv preprint arXiv:2305.18290}.

\bibitem[Rebedea et al.(2023)]{rebedea2023nemo}
Rebedea, T., Dinu, R., Sreedhar, M., Parisien, C., and Cohen, J. (2023).
\newblock NeMo Guardrails: A toolkit for controllable and safe LLM applications with programmable rails.
\newblock \emph{arXiv preprint arXiv:2310.10501}.

\bibitem[Schulman et al.(2017)]{schulman2017proximal}
Schulman, J., Wolski, F., Dhariwal, P., Radford, A., and Klimov, O. (2017).
\newblock Proximal policy optimization algorithms.
\newblock \emph{arXiv preprint arXiv:1707.06347}.

\bibitem[Shao et al.(2025)]{shao2025deepseekmath}
Shao, Z., Luo, Y., Lu, C., Ren, Z.~Z., Hu, J., Ye, T., Gou, Z., Ma, S., and Zhang, X. (2025).
\newblock DeepSeekMath-V2: Towards self-verifiable mathematical reasoning.
\newblock \emph{arXiv preprint arXiv:2511.22570}.

\bibitem[Su et al.(2025)]{su2025crossing}
Su, Y., Yu, D., Song, L., Li, J., Mi, H., Tu, Z., Zhang, M., and Yu, D. (2025).
\newblock Crossing the Reward Bridge: Expanding RL with Verifiable Rewards Across Diverse Domains.
\newblock \emph{arXiv preprint arXiv:2503.23829}.

\bibitem[{Kimi Team} et al.(2025)]{team2025kimi}
Kimi Team, Du, A., Gao, B., et al. (2025).
\newblock Kimi k1.5: Scaling Reinforcement Learning with LLMs.
\newblock \emph{arXiv preprint arXiv:2501.12599}.

\bibitem[Wang et al.(2025)]{wang2025sober}
Wang, Y., et al. (2025).
\newblock A Sober Look at Progress in Language Model Reasoning: Pitfalls and Paths to Reproducibility.
\newblock \emph{arXiv preprint arXiv:2504.07086}.

\bibitem[Yu et al.(2025)]{yu2025dapo}
Yu, Q., et al. (2025).
\newblock DAPO: An Open-Source LLM Reinforcement Learning System at Scale.
\newblock \emph{arXiv preprint arXiv:2503.14476}.

\bibitem[Ziegler et al.(2019)]{ziegler2019fine}
Ziegler, D.~M., Stiennon, N., Wu, J., et al. (2019).
\newblock Fine-tuning language models from human preferences.
\newblock \emph{arXiv preprint arXiv:1909.08593}.

\bibitem[OpenAI(2024)]{openai2024structured}
OpenAI. (2024).
\newblock Introducing Structured Outputs in the {API}.
\newblock \emph{OpenAI Blog}.

\bibitem[Geng et al.(2025)]{geng2025jsonschemabench}
Geng, S., Cooper, H., Moskal, M., Jenkins, S., Berman, J., Ranchin, N., West, R., Horvitz, E., and Nori, H. (2025).
\newblock {JSONSchemaBench}: A Rigorous Benchmark of Structured Outputs for Language Models.
\newblock \emph{arXiv preprint arXiv:2501.10868}.

\bibitem[Dong et al.(2024)]{dong2024xgrammar}
Dong, Y., Ruan, C.~F., Cai, Y., Lai, R., Xu, Z., Zhao, Y., and Chen, T. (2024).
\newblock {XGrammar}: Flexible and Efficient Structured Generation Engine for Large Language Models.
\newblock \emph{arXiv preprint arXiv:2411.15100}.

\bibitem[Shah et al.(2025)]{shah2025rethinking}
Shah, D.~J., Rushton, P., et al. (2025).
\newblock Rethinking Reflection in Pre-Training.
\newblock \emph{arXiv preprint arXiv:2504.04022}.

\bibitem[Lightman et al.(2023)]{lightman2023lets}
Lightman, H., Kosaraju, V., Burda, Y., Edwards, H., Baker, B., Lee, T., Leike, J., Schulman, J., Sutskever, I., and Cobbe, K. (2023).
\newblock Let's Verify Step by Step.
\newblock \emph{arXiv preprint arXiv:2305.20050}.

\end{thebibliography}

\end{document}